# Twins: Device-free Object Tracking using Passive Tags

Jinsong Han[*], Chen Qian[§], Dan Ma[*], Xing Wang[*],

Jizhong Zhao[*], Pengfeng Zhang[*], Wei Xi[*], and Zhiping Jiang[*]

[*]School of Electronic & Information Engineering, Xi'an Jiaotong University

[§]Department of Computer Science, University of Texas at Austin

*Abstract*—Without requiring objects to carry any transceiver, device-free based object tracking provides a promising solution for many localization and tracking systems to monitor non-cooperative objects such as intruders. However, existing device-free solutions mainly use sensors and active RFID tags, which are much more expensive compared to passive tags. In this paper, we propose a novel motion detection and tracking method using passive RFID tags, named Twins. The method leverages a newly observed phenomenon called critical state caused by interference among passive tags. We contribute to both theory and practice of such phenomenon by presenting a new interference model that perfectly explains this phenomenon and using extensive experiments to validate it. We design a practical Twins based intrusion detection scheme and implement a real prototype with commercial off-the-shelf reader and tags. The results show that Twins is effective in detecting the moving object, with low location error of 0.75m in average.

## I. INTRODUCTION

Preventing illegal or unauthorized access by intruders is of importance to protect the security of people and their properties. For anti-intrusion, there is an essential need to deploy security systems well-supported by localization and motion-detecting methods. To this end, existing work performs motion detection using various sensors including passive infrared (PIR) sensors, sonic sensors, and video camera sensors. Those methods, though being able to achieve high accuracy and sensitivity, are not cost-efficient for large scale logistic systems, such as retailing, warehouse, cargo transportation, due to various reasons. In recent decades, Radio Frequency Identification (RFID) tags have been widely deployed in modern logistic and inventory systems for efficient identification and monitoring. Compared with deploying sensor systems, motion detection using RFID tags for anti-intrusion purpose has two main advantages: low cost and reuse of existing RFID infrastructure.

Motion detection or trajectory tracking by RFID tags has been proposed in the literature [1-7]. There are two major categories in previous work, device-based and device-free methods. By attaching a tag to an item, the device-based method can identify the location of this item when the tag is interrogated by the reader. However, it is impossible to bind tags to uncooperative objects, e.g. the intruders. Thus device-based methods are not suitable for intrusion detection and tracking in many practical applications. Device-free solutions are promising to track intruders while keeping them unaware of detection [8].

However, most existing solutions are based on active tags to achieve device-free object tracking [3, 7, 8, 10]. An active tag is much more expensive than a passive tag, which is too cost-ineffective for large-scale applications. To our knowledge, there is no solution in the literature that can achieve accurate and reliable device-free motion detection with passive RFID tags.

In this paper, we present a novel motion detection and tracking method using passive RFID tags, named *Twins*. Our method is motivated from a newly observed phenomenon due to the coupling effect among passive tags. Suppose we put the antennas of two passive tags close and parallel to one another, as illustrated in Fig.1. Within a certain distance, the two adjacent tags will present such a phenomenon that one of them, e.g. tag *B* in Fig.1, JUST becomes unreadable, due to the coupling effect. It is like that tag *A* casts a "shadow" to tag *B*. As a result, the signal strength of RF waves sent from the reader is significantly reduced at *B*'s antenna. As a result, tag *B* cannot receive sufficient energy to perform the computation or modulation, and cannot be read by the reader. We name such a situation as *critical state* and such a pair of tags as *Twins* or *a Twin pair* here after.

We then utilize this phenomenon to achieve device-free motion detection. The key insight is to create a critical state of two tags as Twins. If an object or human being moves close to the Twins, as shown in Fig.1, some RF waves will be reflected or refracted to the Twins, similar to the multipath effect. In this case, the unreadable tag can receive sufficient energy to break the critical state and then become readable. We call this change as a *state jumping*.

However, existing interference models of RFID tags contradict to our observation from real experiments and cannot explain the critical state phenomenon of Twins. It is because they are "structure-oblivious" models without considering the structure of antenna circuit.

We develop a novel "structure-aware" interference model of RFID tags based on the analysis of the circuit structure, T-match [12], which plays an important role in the interaction between two adjacent tags. Comprehensive experimental results agree the proposed model and validate the occurrence of the critical state.

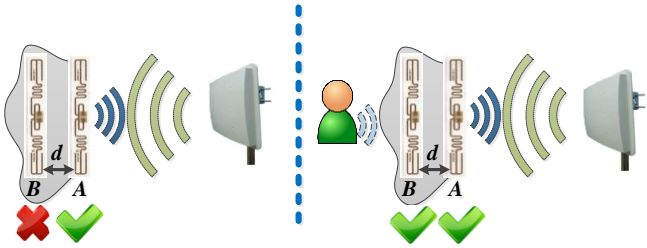

Fig. 1. Critical state of Twins.

We further design a tracking scheme to achieve accurate trajectory monitoring for moving objects. Our scheme employs a combination of a KNN based algorithm and particle filter based algorithm to approximate the trajectory of intruders. We summarize the major contributions and results of our work as below.

1) We are the first to propose to use the critical state of passive tags for intrusion detection and trajectory tracking. By reusing the existing RFID infrastructure, this method is a cost-effective and accurate anti-intrusion solution.

2) We contribute to the theory of the tag interference model and develop a "structure-aware" model which perfectly explains the critical state phenomenon.

3) We conduct extensive experiments on a real RFID system for validating the feasibility of using critical states for motion detection.

4) We design a tracking scheme for effective intruder detection and tracking. We have implemented it in a real RFID system using off-the-shelf reader and tags. The experimental results show that our solution can achieve high detection accuracy, i.e. the localization error is 0.75m in average.

## II. BACKGROUND

As the most representative technology of "untouchable" identification, RFID shows many advantages over the conventional labeling techniques. RFID tags can be automatically and remotely interrogated in a non-line-of-sight way. Current RFID tags fall into two categories, active and passive tags. With on board batteries, active tags have a larger communication region and more powerful computation capacity than passive tags, while suffering from much higher manufacture cost. Passive tags are more cost-efficient. They are battery-free and with simple circuitries.

### A. Near-field and Far-field

In RFID systems, the reader and tags communicate via their antennas. The communication patterns include near-field and far-field communications. The boundary between near-field and far-field is determined by the Rayleigh length [13], calculated as $R = 2D^2/\lambda$, where $D$ is the size of antenna and $\lambda$ is the wavelength of antenna. Equivalently, $D$ is the diameter of the smallest sphere enclosing the antenna. Passive tags are identified by using the far-field communication. According to FCC regulation, passive tags should operate in a spectrum of 902~928 MHz in US. With far-field links, RFID reader interrogates the tag by emitting RF waves, while the passive tag modulates its data into the wave reflected to the reader. This pattern is referred as *Backscatter* communication [12]. On the other hand, the interference between two adjacent tags takes effect as a *coupling effect* in the near-field.

### B. Dipoles and T-match structure

Most passive tags use a half-wave dipole antenna. We show a commercial passive tag, Impinj E-41b in Fig.2. The length of antenna should be $\lambda/2$, i.e. 16cm for 915MHz. To reduce the physical size, the antenna is bent to form a *meandered dipole*. However, meandered dipole faces a problem of matching. A large mismatch of antenna to IC may result in a small power transfer coefficient, and hence a small portion of received power to be used by the tag. One practical treatment for improving the match is to adapt a short antenna to the capacitive IC load, forming a T-match structure, as shown in Fig.3. In this structure, the impedance of the longer meandered dipole (with the length of *L*) can be tuned by the introduced shorter dipole (with the length of *a*). The IC of the tag connects to the meandered diploe via two wings of the short dipole.

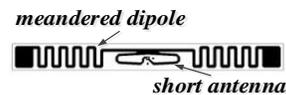 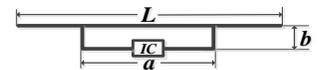

Fig.2. Real tag model.    Fig.3. T-match structure.

## III. CRITICAL STATE OF COUPLING RFID TAGS

In this section, we first show the conflict between the conclusion derived from conventional "structure-oblivious" model and the result observed from real experiments. To address the contradiction, we propose a hypothesis that cannot be explained by the conventional method. We then use a "structure-aware" model for explaining the interaction of Twins. The model is theoretically proved and perfectly matching the experiment result.

### A. Obeservation from experiments of Adjacent Tags

We conduct a set of experiments using 20 off-the-shelf tags (Impinj E41-b). We randomly label these 20 tags by $A_1$, $A_2$, …, $A_{10}$, $B_1$, $B_2$, …, $B_{10}$ and then form 10 sets of Twins $(A_1, B_1)$, …, $(A_{10}, B_{10})$. We place the Twins in a plastic foam board. For different pairs of Twins, we change the placement of two tags and record the minimum power needed for reading them, as shown in Fig.4. In these experiments, the distance between two tags is fixed to 10mm, and the distance between tags and the reader's antenna is fixed to 2m.

Since increasing the transmission power of the reader can increase the coupled current in tags, *the minimum power value reflects the scale of current to activate a tag*. For each sub-figure, we simplify the geometry shape of tags for illustrating their relative positions. We define the tag with its IC as the closest part to another tag than its other parts as the *Rear*-tag in Twins, like Tag A in Fig.4(a). Correspondingly, Tag B will be the *Fore*-tag in Fig.4(a). The average minimum powers for each tag are shown with bar chart of Fig.4. In Fig.4(a) – (d), the difference of minimum power in a Twins is about 10dbm. In other deployments (Fig.4(e) - (h)), the two tags' minimum powers are almost the same. The big difference of minimum power in Fig.4(a) - (d) means there exists the "shadowing" effect that causes critical state for Twins. In addition, we can find that in Twins, the Rear-tag is the "shadowed" one that needs more power to be activated.

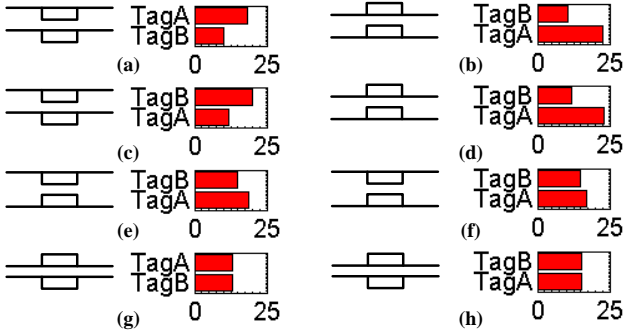

Fig. 4. The placements of Twins and the minimum $P_{TX}$ used for reading them.

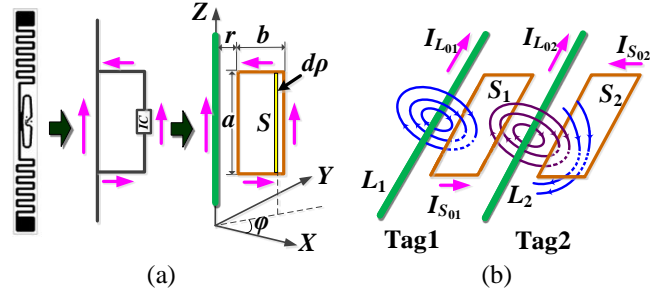

Fig. 5. Structure-aware modeling for the Twins.

However, none of existing RFID interference models [14, 15] can explain the existence of the shadowing effect. Based on these models, the two adjacent tags are modeled as two identical circular loops. Since the two tags are equal in the distance to the reader and deflection angle from the axis of the reader's antenna in our experiments, their currents coupled by the reader's RF signals, $I_{01}$ and $I_{02}$ should be equal $I_{01} = I_{02}$. Based on the *Reciprocity Theorem* which combines the Ampere's law and Biot-Savart law [13], the two mutual inductances are equal, $M = M_{12} = M_{21}$. For the current in Tag 1, we have $I_1 = I_{01} - bM_{12}I_2$, where $b$ is an factor representing the equivalent RCL circuit of the Tag1, which is also equal to that of Tag2. Similarly, we have $I_2 = I_{02} - bM_{21}I_1$. Obviously, $I_1 = I_2$. Here $I_1$ or $I_2$ is integrated result of current coupled by the reader's RF signals and the mutual coupling between Tag1 and Tag2. (detailed derivation can be found in our technical report [9]).

Such a result obviously contradicts our observation in the experiments. It is because these models are "structure-oblivious", meaning they do not consider detailed antenna structure that may cause different interference levels for different placements.

### B. Structure-aware model

We introduce a "structure-aware" interference model that can provide reasonable explanation to the phenomenon of shadowing. Instead of using a simply circular loop, we use two dipoles to model the antenna of passive tags, as shown in Fig.5(a). Note Tag1 is the Rear-tag. We model the dipole antenna as two components as shown in Fig.5(a). One is an electric dipole like a line and another is a magnetic dipole like a rectangle. This modeling is well coincident with the T-match structure.

For Twins, we get four conductors, two lines ($L_1, L_2$) and two rectangles ($S_1, S_2$), as shown in Fig.5(b). Suppose the coupling currents contributed by the reader's RF signals in the two lines are $I_{L_{01}}$ and $I_{L_{02}}$, respectively. As the tag's IC is embedded into the magnetic dipole, the current in $S_1$ contributes most to the energy activating the tag, i.e. $I_1 = I_{S_1}$. Similarly, $I_2 = I_{S_2}$. We have the following theorem.

*Theorem:* If two tags *Tag1* and *Tag2* are placed as in Fig.5(b), let $I_{S_1}$ and $I_{S_2}$ be the current on $S_1$ and $S_2$, respectively. We have $I_{S_1} < I_{S_2}$

*Proof:* As previously indicated in Section III-A, we have $I_{L_{01}} = I_{L_{02}} = I_{L_0}e^{j\omega t}$, where $I_{L_0}$ is the complex amplitude. The area of the rectangle is $a \times b$. Let $r$ denote the distance from the rectangle to the line ($r$ is very slight). We assume the $I_{L_{01}}$ and $I_{L_{02}}$ are in the same direction as Z axis. We analyze the interaction among these four conductors as follows.

**1) Mutual Inductance in $S_1$**

The mutual inductance in $S_1$ is an integration of the mutual inductances between $L_1$ and $S_1$, between $L_2$ and $S_1$, and between $S_1$ and $S_2$.

**a).** Mutual inductance between $L_1$ and $S_1$

The magnetic induction $B$ yielded by current $I_{01}$ can be represented as concentric circles around $L_1$. In the cylindrical model in Fig.5(a), the magnetic induction $B$ in position ($\rho, \varphi, z$) can be calculated by using the *Biot-Savart* Law [13] as

$$\vec{B} = \vec{e_\varphi} \frac{\mu_0 I_{L_{01}}}{2\pi\rho} \quad (1)$$

where $\mu_0$ is the magnetic constant and $\vec{e_\varphi}$ is the unit direction vector of the magnetic induction.

The magnetic flux caused by $I_{L_{01}}$ within the rectangle $S_1$ can be represented as

$$\Psi_{11} = \int_{S_1} \vec{B}_{S_1} d\vec{S} \quad (2)$$

We denote $d\rho$ as the tiny rectangle facet primitive within the $S_1$, and then $d\vec{S} = ad\rho$, where $a$ is the width of the rectangle. Calculating the integral from $r$ to $r+b$, where $b$ is the rectangle's length, we have

$$\Psi_{11} = \frac{\mu_0 I_{L_{01}} a}{2\pi} \int_r^{r+b} \frac{1}{\rho} d\rho = \frac{\mu_0 I_{L_{01}} a}{2\pi} \ln\left(\frac{r+b}{r}\right) \quad (3)$$

As discussed in Equation 2 in our technical report [9], the mutual inductance $M_{11}$ of $L_1$ and $S_1$ is

$$M_{11} = \frac{\Psi_{11}}{I_{L_{01}}} = \frac{\mu_0 a}{2\pi} \ln\left(\frac{r+b}{r}\right) \quad (4)$$

The induced electromotive force in $S_1$ is

$$\Delta E = -\frac{d\Psi_{11}}{dt} = -\frac{M_{11} dI_{L_{01}}}{dt} = -j\omega I_{L_0} e^{j\omega t} M_{11}$$
$$= -\frac{\mu_0 a j\omega I_{L_0} e^{j\omega t}}{2\pi} \ln\left(\frac{r+b}{r}\right) \quad (5)$$

Since the magnetic flux of $L_1$ is in an opposite direction with that of $S_1$, the two opposing magnetic fluxes have a force of canceling out to each other. Thus, the induced electromotive force in $S_1$ is negative: $E_{11} = -\Delta E$.

Suppose the equivalent resistance in $S_1$ and $S_2$ is $R$, the current coupled by $I_{L_{01}}$ in $S_1$ is

$$I_{11} = \frac{E_{11}}{R} = \frac{\mu_0 a j\omega I_{L_0} e^{j\omega t}}{2\pi R} \ln\left(\frac{r+b}{r}\right) \quad (6)$$

**b).** Mutual inductance between $L_2$ and $S_1$

The coupling effect between $L_2$ and $S_1$ can be derived similarly. The mutual inductance can be written as

$$M_{21} = \frac{\Psi_{21}}{I_{L_{01}}} = \frac{\mu_0 a}{2\pi} \ln\left(\frac{l+b}{l}\right) \quad (7)$$

On the other hand, the direction of magnetic flux of $I_{L_{02}}$ is the same as that of $S_1$, the integrated magnetic flux is enhanced. We have

$$I_{21} = -\frac{\mu_0 a j\omega I_{L_0} e^{j\omega t}}{2\pi R} \ln\left(\frac{l+b}{l}\right) \quad (8)$$

**c).** Mutual inductance between $S_1$ and $S_2$

According to *Reciprocity Theorems* [13], $S_1$ and $S_2$ is symmetric to each other. The coupled current, denoted as $-I_H$, has an identical value in $S_1$ and $S_2$, while the coupled current is always in the opposite direction to the exciting current.

**2) Mutual Inductance in $S_2$**

Similar to Equation 6 and 8, we can calculate the coupling effect between $L_1$ and $S_2$, and the coupling effect between $L_2$ and $S_2$, which are denoted as $I_{12}$ and $I_{22}$.

$$I_{12} = \frac{\mu_0 a j\omega I_{L_0} e^{j\omega t}}{2\pi R} \ln\left(\frac{2r+2b+l}{2r+b+l}\right) \quad (9)$$

$$I_{22} = \frac{\mu_0 a j\omega I_{L_0} e^{j\omega t}}{2\pi R} \ln\left(\frac{r+b}{r}\right) \quad (10)$$

where $l$ represents the distance between $S_1$ and $L_2$, and $l \approx d$.

**3) Currents in $S_1$ and $S_2$**

Suppose $I_{S_{01}}$ and $I_{S_{02}}$ are the currents in $S_1$ and $S_2$ generated by harvesting the RF signals from the reader's antenna. Then we can get the representations of currents in $S_1$ and $S_2$:

$$I_{S_1} = I_{S_{01}} + I_{11} + I_{21} - I_H$$
$$= I_{S_{01}} - I_H + \frac{\mu_0 a j\omega I_{L_0} e^{j\omega t}}{2\pi R} \ln\left(\frac{r+b}{r}\right) - \frac{\mu_0 a j\omega I_{L_0} e^{j\omega t}}{2\pi R} \ln\left(\frac{l+b}{l}\right)$$

$$I_{S_2} = I_{S_{02}} + I_{12} + I_{22} - I_H$$
$$= I_{S_{02}} - I_H + \frac{\mu_0 a j\omega I_{L_0} e^{j\omega t}}{2\pi R} \ln\left(\frac{2r+2b+l}{2r+b+l}\right)$$
$$+ \frac{\mu_0 a j\omega I_{L_0} e^{j\omega t}}{2\pi R} \ln\left(\frac{r+b}{r}\right) \quad (11)$$

We know that $I_{S_{01}} = I_{S_{02}}$, from the analysis using the "structure-oblivious" model. Thus, $I_{S_2} > I_{S_1}$. ∎

It is obvious that $l$ and $b$ are crucial in the critical state. When they are in a same length scale, i.e., the two tags are adjacent, the critical state can be easily triggered, due to a large difference between $I_{S_1}$ and $I_{S_2}$. In particular, $I_{S_2} > I_{S_1}$ explicitly explain why the Rear-tag always becomes unreadable in our experiments. We can further calculate the critical condition to make the inductive coupling effect occur for Twins. From Equation 11, when $l \gg b$, we can get

$$\lim_{\frac{b}{l} \to 0} \frac{l+b}{l} = 1 \quad \lim_{\frac{b}{l} \to 0} \frac{2r+2b+l}{2r+b+l} = 1$$

$$\lim_{\frac{b}{l} \to 0} \ln\left(\frac{l+b}{l}\right) = 0 \quad \lim_{\frac{b}{l} \to 0} \ln\left(\frac{2r+2b+l}{2r+b+l}\right) = 0$$

and $I_{21} = I_{22}$.

In this case, $I_{S_{01}} \approx I_{S_{02}}$, which means the inductive coupling effect of the two tags is nearly identical to each other, when the distance between two tags is sufficiently large.

## IV. VALIDATION OF THE CRITICAL STATE

In this section, we experimentally validate the critical state and state jumping phenomenon. First, we deploy two tags A and B as Twins and a reader as shown in Fig.6. We then create a critical state of the Twins by tuning the transmission power of the reader, $P_{TX}$. For enabling a fine-grained tuning, we employ a UI of Impinj readers, MultiReader. In this phase, we first set $P_{TX}$ to 32.5 dBm, and then gradually decrease $P_{TX}$. For each step, the reader attempts to interrogate the Twins. A critical state occurs if one tag becomes unreadable.

Let $d$ denote the distance between two tags, and $D$ denote the distance from the reader to the Twins. We vary $d$ from 6mm to 26mm and try to yield a critical state for the Twins with a fixed $D$ of 2m. We find that a critical state occurs when the distance is less than 15mm as shown in Fig.7. Tag B, which is the Rear-tag in the Twins, is always the one that cannot be read in the critical state. Tag B shows a higher minimum value of $P_{TX}$ than that of Tag A. When $d$ decreases, the difference between values of $P_{TX}$ of the two tags becomes larger.

In the next phase, a volunteer moves around the Twins for investigating the region where the critical state works for motion detection. We record the occurrence of state jumping caused by the moving volunteer and show the corresponding positions in Fig.8. We divide the experiment area into small cells, and count the number of state jumping events in each cell.

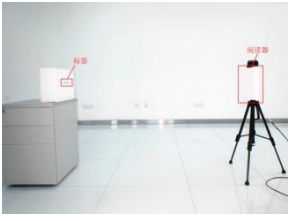
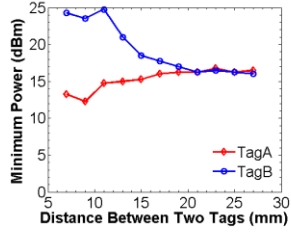

Fig. 6. Validation setup.   Fig. 7. Minimum $P_{TX}$ vs. $d$.

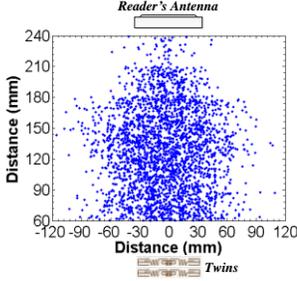
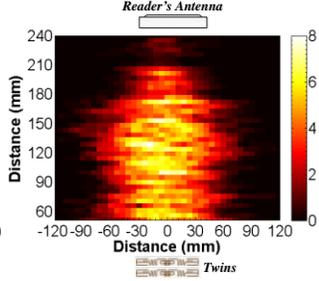

Fig. 8. # of state jumping.   Fig. 9. Effective region.

The results are plotted in Fig.9. This figure geometrically illustrates the sensitivity of detecting the moving object. It can be found that the effective range of detection is relatively large, about a 2m × 1m rectangle between the reader and Twins. When the volunteer moves behind the Twins, the effective range of detection is reduced to 1m away from the Twins.

## V. TRACKING MOVING OBJECTS

The tracking procedure comprises of two phases. In the first phase, the state jumping on Twins are identified. In the second phase, we use a particle filter [16] based scheme to track the object. Note that we focus on tracking a single object in this paper and leave the multiple-object tracking task in future work.

A number of Twins are deployed in the given area, for instance, the alleyway between two shelves with valuable items. We assume each Twins is fully covered by at least one reader's interrogation. The entire region is partitioned into a 2D grid, as shown in Fig.10. In practice, the distance between two Twins, i.e., the length of cells edges, can be determined based on measurement results. We can formulate the grid as a graph $G = <V, E>$, where each cell in the grid is a vertex in $G$, and two adjacent cells have an edge in $G$. If state jumping is detected at Twin pairs, the corresponding cells will be highlighted as shown in Fig.10.

### A. Identifying state jumping

When an object moves in the monitored region, it triggers a serial of state jumping on multiple Twin pairs. Timely outlining the regions including these Twin pairs is essential to track the object movement. However, there remains a challenging issue. At any time, the reader can only have a fixed value of transmission power $P_{TX}$. Therefore the reader can only create the critical state for one Twin pair at a same time. Intuitively, the reader can iteratively and sequentially query all Twin pairs within its detection region. However, sequentially query may miss state jumping, if the movement is so fast that the object has moved out of the cells when the Twin pairs are queried. Using a more sophisticated strategy, the reader can preferentially query the nearby Twin pairs of the Twins that just had state jumping. However, scheduling starvation may occur on some Twin pairs if they have low priority in such scheduling.

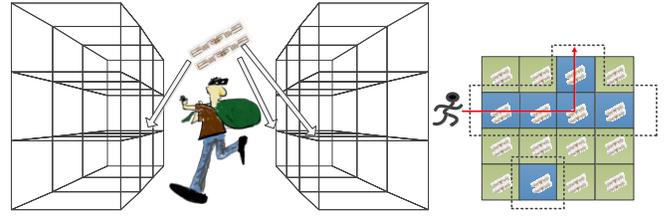

Fig. 10. Intruder detection in warehouse.

To deal with this problem, we propose a simple polling algorithm based on the graph $G$ to achieve timely and efficient movement detection without introducing starvation. Suppose the system has $N$ Twin pairs. The objective of the polling algorithm is to find a set $J$ containing the Twin pairs with state jumping caused by object movement within a short time interval $\Delta t$. The polling algorithm is Multi-level Priority Linked-List based (MPLL). We first construct two linked lists $L_P$ and $L_N$. $L_P$ is with higher priority than $L_N$, and used for containing the tuples of Twins pairs experiencing state jumping in last polling round. For the $i$th Twin pair, we store the tuple $t = <P_{TX, i}, T_i, P_i, S_i>$ in the database, where $T_i$ denotes the sequence number of the Twins, $P_{TX, i}$ denotes the corresponding transmission power of reader's antenna to create the critical state of $T_i$. $P_i$ and $S_i$ are two bits representing the query priority and access status of $T_i$. If a Twins pair's $P_i$ is 1, it will be put in $L_P$, otherwise in $L_N$.

The basic idea behind MPLL is as follows. We allow a reader iteratively query the Twins pairs on the $G$. The query is performed in a Breadth-First Search (BFS) pattern. For a Twins pair upon the query, the reader will check whether it is experiencing a state jumping in this round. If the answer is Yes and this Twins pair is in $L_P$, the reader keeps the Twins pair in $L_P$, and then queries all its neighboring Twins pairs in $G$ via BFS. This treatment is for acquiring as many Twins pairs that are triggered by the object movement as possible. If the answer is No, the reader will move this Twins pair to the end of $L_N$. After the query on current Twins, the reader moves to its successor in $L_P$. Note the BFS is recursively executed in $G$. For a Twins pair in $L_N$ experiencing a state jumping, the reader will set its priority bit as 1 and simply move it to the end of $L_P$, otherwise move to its successor Twins pair in $L_N$.

Some Twin pairs that are likely to have state jumping may be queried by multiple times in one polling round. We use the $S_i$ to indicate whether the Twins pair has been accessed in a polling round and hence avoid unnecessary queries. In short, the algorithm can identify all Twin pair having state jumping in a short time interval with high probability. Due to the limitation of space, the detail of MPLL algorithm can be found in our technical report [9].

### B. Localizing an object

Based on the detected state jumping on Twins, we can outline the region where the object stays. A smaller region results in a more accurate location. Ideally, the region should be represented by a connected subgraph $G_s$ in $G$. However, the movement of an object may render several subgraphs unconnected in $G$, due to the multi-path effect or ambient noise. We show an example of two separate regions in Fig.11. In this case, we simply select the largest subgraph in $G$, termed as $G_s$ and filter other disconnected ones out. If there are more than two subgraphs, $G_s$ is set to the minimum connected component in G that includes all subgraphs.

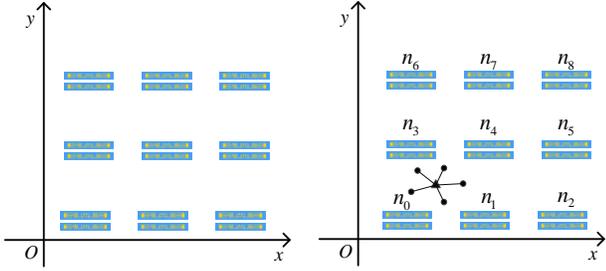

(a) Initial status of the Twins grid  (b) an object is in the area

Fig.11. Particle filter.

After determining the possible region that the object stays, we estimate the current position of the object by calculating the centroid of the positions of all Twins in this region. Suppose there are $k$ Twin pairs in the region whose positions are $X_1, X_2,..., X_k$, respectively. The estimated position is C = $(X_1+ X_2+\cdots+X_k )/k$. This treatment might increase detection errors. We check the tracking accuracy in Section VII.

### C. Tracking the object based on the particle filter

A particle filter is designed to optimize the estimation for non-linear and non-Gaussian state-space models. The main principle of particle filter is to introduce a group of "particles", which actually are random samples in the state space. Utilizing those particles, the distribution of a latent variable can be "filtered" (approximated) at a specific time, given all observations up to that time. By iteratively filtering and re-sampling, the state of the system or targeted object can be estimated. If the state contains the position or derivatives of position, we can achieve the trace of the object.

Suppose the object moves into the monitored area. We set the main entrance of the area as the origin of coordinates. In Fig.11, $n_0$~$n_8$ represent the times of state jumping at the corresponding Twins in $\Delta t$. We define the vector $V = (n_0, n_1, n_2, \ldots, n_8)$ as the *observation*.

The particle filter is performed via two phase, offline training and online detecting. In the former phase, we estimate the marginal distribution of $n_i$ and the position of object $l$ $P(n_1,n_2,\ldots,n_N|l) = \prod_{i=1}^{N}P(n_i | l)$. Note that $l$ is known in the offline training phase. The estimation discretizes the distribution $P$ into histogram bins, and forms the fingerprint of the position $l$.

In the second phase, the particle filter is conducted by performing the following steps.

a. Initialization. Suppose the object is located in the origin of coordinates, with a speed as $v = <x_v, y_v>$, where $x_v$ and $y_v$ are the X-component and Y-component of speed $v$, respectively. $N$ particles are sampled near the origin of coordinates.

b. Prediction. The $p(X_k|Z_{1:k-1})$ is computed from the filtering distribution $p(X_{k-1}|Z_{1:k-1})$ at time $k$-1.

c. Weight computation. The position of each particle is known. Computing the weight depends on two elements, the observation obtained by the measurement as well as the probability distribution of the times of state jumping at each particle's position. The weight is indeed the probability of yielding the observation at each particle.

d. Re-sampling. We use the multinomial resampling [17] to remove the particle with low weight with high probability.

e. Approximation. With the position and weight of each particle, we can approximate the coordinates for the object at time $k$.

f. Let $k = k$ +1, and iteratively perform the steps (b) – (f).

For the detail procedure of particle filter based tracking algorithm, please refer to our technical report [9].

## VI. DISCUSSION

### A. Proper detection region

Observed from experiments, state jumping is much easier to occur if the object sits between the Twins and the reader, instead of being behind the Twins. It is mainly because the object moving behind triggers a weaker disturbance to the RF signals around the Twins than in the "in-between" space. Furthermore, the detection becomes much unreliable if the Twins is attached to items. In our implementation and experiments, we adopt the "in-between" deployment for the reader and Twins.

### B. Critical state of a single tag

Instead of using two adjacent tags, we can use a single tag for the moving detection via its own critical state. That is, the reader can probe using the minimum transmission power that can identify a tag in a given position. If one object moves nearby, the tag may also become readable. We perform the experiment on single tag and find that the single tag is, however, more unstable than Twins in the generation and maintenance of critical states. The reason is that the single tag is more susceptible to the ambient changes. While this may lead more sensitivity to the single tag, the stability, i.e., $P_{TX}$ required at the reader for generating critical state varies much drastically than Twins. We leave the study of addressing the stability of critical state for the single tag in our future work.

### C. Multiple moving objects

In this work, we focus on the tracking of single object. If the area monitored has multiple moving objects, can we still be able to detect them based on Twins? Indeed, the Twins based tracking scheme has a potential solution. Actually, if there are multiple objects moving in the area, their movements may result multiple subgraphs in $G$. Intuitively, we can track those objects via those subgraphs. However, there are many challenges in plotting the subsequential traces for those objects, considering the scenario that their moving trajectories may cross or overlap. Since tracking multiple objects is out of the scope of this paper, we will give a solution in our future work.

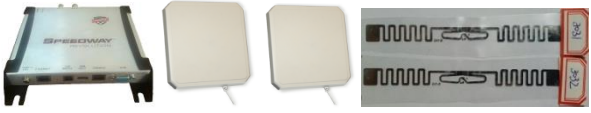

Fig. 12. Experiment hardware

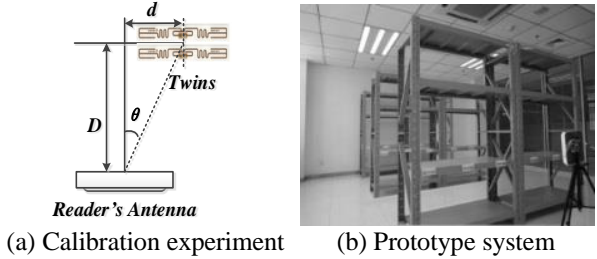

(a) Calibration experiment    (b) Prototype system

Fig.13. Experiment setup.

*D. System overhead*

The cost would be a major consideration on this work, due to the relatively expensive price of readers and antennas. If we can only afford a few of RFID readers, there will be many uncovered regions if the surveillance area is large. As a consequence, the system may fail to effectively detect the moving object. We argue that the work is suitable for the logistics scenarios where an RFID infrastructure has been available so that the critical areas will be covered by RFID readers. Nevertheless, the cost is still a challenge. In this case, we can deploy the reader and antenna to cover crucial regions, such as the major entrances or shelves containing valuable items. Furthermore, the problem of tracking an object is to identify the sequence of its path segments that is most likely to produce the observed sequence of Twins experiencing a state jumping. There are many effective approaches available to solve above problem, such as the HMM and Viterbi decoding used in [2]. We will address this problem in our ongoing work.

## VII. IMPLEMENTATION AND EVALUATION

In this section, we describe the prototype implementation and performance evaluation. We conduct the experiments in two aspects. First, we investigate the performance of critical state creation and detection. Second, we implement a Twins prototype and conduct comprehensive experiments to evaluate the effectiveness and efficiency of object tracking.

*A. Hardware*

Figure 12 shows the key hardware components in our prototype. The hardware is entirely built from current commercial products and requires no modification on both the reader and tag sides. The reader is a Impinj Speedway R420, using the EPCglobal UHF Class 1 Gen 2/ISO 18000-6C air protocol. The reader antenna operate in a spectrum of 920~928MHz. The transmission power ranges from 10 to 32.5 dBm. We use 500 E41-b tags in our experiments. E41-b is a widely-deployed off-the-shelf passive tag from Impinj.

*B. Methodology*

  1) *Calibration*

The goal of calibration is to determine the proper settings of Twins for motion detection. We investigate the performance of one Twins in detecting moving objects nearby. As shown in Fig.13(a), we find proper values for distance $d$ and angle $\theta$ between the antennas of Twins and the reader, distance between the Twins and moving object, and height from the Twins to the floor $h$.

  2) *Prototype*

In the experiment, we implement our prototype system for the surveillance in a warehouse. We setup a testbed with a number of real shelves aligned as shown in Fig.13(b). Each Twins is attached to the shelf. The distance between two shelves is 2m and the distance of two adjacent Twin pairs is 0.6m. A volunteer walks among the shelves. We examine the performance of detection rate and probe the proper settings for practical deployment.

  3) *Experiment environment*

We run the experiments in a large warehouse to check the tracking accuracy of Twins. The area is 30m×20m and deployed with Twins grid. Height of the intruder is 1.70m and the moving speed is 1.5m/s.

*C. Performance Evaluation*

  1) *Key parameters of Twins*

Through extensive calibration, we investigate the crucial settings in the creation of critical state. We evaluate the impact of different factors on the successful detection rate $r$ of the moving object in the monitoring region.

The main lobe width of the reader's antenna used in the prototype is 70°. Thus, it is not necessary to set a $\theta$ larger than 35°. We exam the value of $r$ by varying $\theta$ as 0°, 15° and 30°. For each $\theta$, we vary the distance $D$ to 75cm, 105cm, 135cm, and 165cm. With each distance $D$, we mutually set the distance $d$ as 6mm, 8mm, 10mm, 12mm, and 15mm. Therefore, we have 3*4*5=60 test cases. For each case, we conduct 100 runs of test. The result is summarized in Fig.14. The X axis contains different settings of $d$. The Y axis represents the $r$. We differentiate the settings of angle with different colors. We have an observation that setting $d = 10$mm can result in the maximum $r$ in most cases. We then adopt a default setting of $d$ as 10mm in the following experiments.

We invite three volunteers to play the role of intruders and walk among the shelves in our prototype. The three volunteers are 160cm, 170cm, and 180cm in height, respectively. Their average walking speed is 1.5m/s. The result is reported in Fig.15. We find that the system has a higher detection rate on taller persons. The reason is that the taller person has a larger cross-sectional area such that more RF waves can be reflected to the Twins, yielding a state jumping with higher probability. We also notice that the detection rate is always over 80% for all volunteers, and 90% for the volunteers with a 1.70m+ height. This result show that our system is relatively height-insensitive in detecting human movements.

Our system is also feasible when changing the distance between Twins and reader antenna $D$. We vary the value of $D$

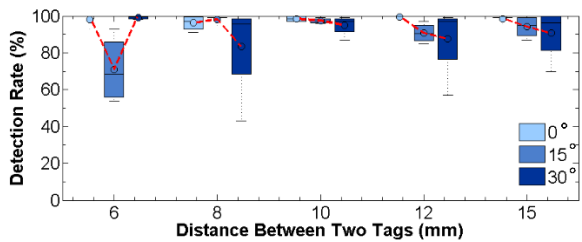

Fig. 14. Calibration result for single Twins.

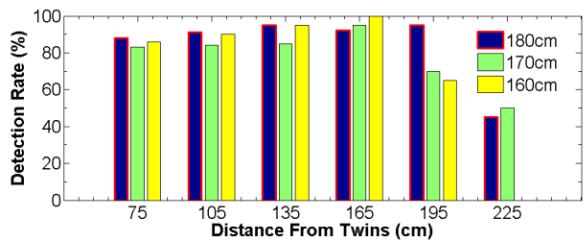

Fig. 15. Impact from the object with different heights.

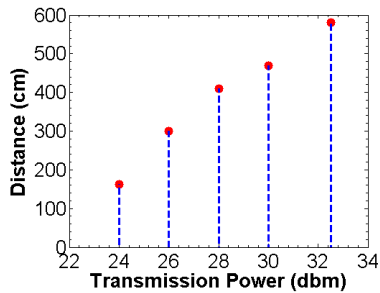

Fig.16. $P_{TX}$ vs. $D$.

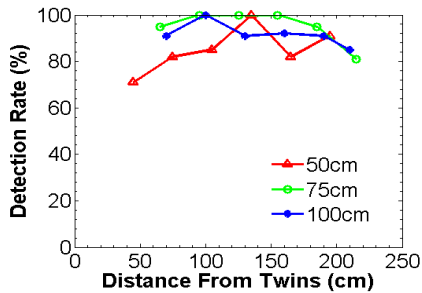

Fig.17. The height above the floor.

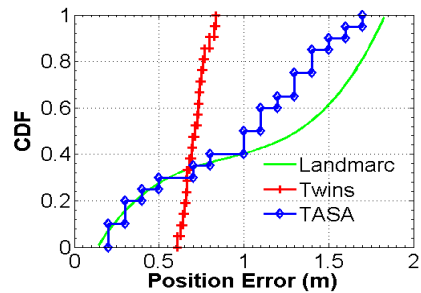

Fig. 18. Tracking accuracy.

and exam the transmission power $P_{TX}$ to drive the Twins into the critical state. As the result shown in Fig.16, we find that a larger $D$ requires a higher $P_{TX}$. Therefore, the maximum deployment distance of Twins depends on the maximum transmission power of reader, e.g. 5.8m when $P_{TX} = 32.5dbm$ if using the Impinj R420 reader.

We then probe a proper setting on the height of Twins above the floor in real deployment. Let $h$ to denote this value. Figure 17 shows the successful detection rate with different settings of $h$. The X axis in Fig.17 represents the distance between the volunteer and Twins. We notice that when $h$ equals to 75cm, Twins can achieve the highest detection rate (95.17% in average). This is because when $h$=75cm, the movement of arms and legs helps to reflect more RF signals to the Twins via multipath effects, which is easier to trigger a state jumping. However, when the height of Twins is too low or too high (e.g. 50cm or 100cm), the movement leads a smaller impact on the Twins, and results in a lower detection rate.

**2)** *Tracking accuracy*

We investigate the accuracy of Twins-based tracking scheme and compare it with two well-known RFID based device-free approaches, LANDMARC [3] and TASA [7] through experiments. LANDMARC is active tag based and TASA is a hybrid systems of active and passive tags. The tags are deployed in a tag array in all the three approaches. The distance between the nearest neighbors in a row or column is 1m for LANDMARC and 0.6m for TASA. During the procedure of tracking the simulated intruder, we record the distance from the estimated position to the real position as the localization error for each detection. The result plotted in Fig.18 exhibits that Twins based scheme has a better tracking accuracy than both of LANDMARC and TASA in average. Although Twins is not accurate in a small portion of positions on the moving path, its position error rate is always below 0.85m, and 0.75 in average.

## VIII. RELATED WORKS

The idea of utilizing wireless signals for activity sensing is not new [3]. It continuously attracts attentions recently, due to the prevalence of today's wireless and mobile devices. Research community has proposed many proposals for localizing and tracking objects by using rich sensors and various context attributes, including the GSM [18], WiFi [19], GPS [20], FM [21], and acoustic signal [22]. From the perspective of tracking pattern, prior works in the literature can be categorized into two groups, device-based and device-free approaches. The former pattern normally requires a device to be bound with the target, while the latter one has no need to bind a device to the target.

Device-based approaches require the target attaching or holding a tracking device. Among those works, LANDMARC [3] is a pioneering work of exploiting the RSS change for localization and tracking. It first site-surveys the RSS-based position fingerprint of a tag-array-covered area. Tracking an active tag can be realized by matching the collected RSS change with the position fingerprint [3] [10]. There is a growing interest in crowdsourcing the sensing capacity of a large amount of computing devices that are held by unprofessional users, such as the works Zee [23] and LiFS [24].

Passive tag based localization is often deployed in the warehouse or library for accurately locating the desired item or book [4, 11, 25]. Choi et al [25] propose a localization algorithm LDTI. LDTI locates the box in a shelf by detecting the tag-interference among the reference tags and target tag. PinIt [11] is one of the most recent works that exploit a passive tag's multipath profile for positioning the object. The work employs the synthetic aperture radar (SAR) imaging mechanism to achieve high accuracy in locating the passive tag. The work shows a promising direction for leveraging multipath effect for localizing low-cost mobile devices.

However, the device-based works requires the desired object to be with a device for localization, which are orthogonal to Twins.

On the other hand, the Device-free Passive (DfP) localization is more suitable for monitoring uncooperative objects. Most prior works leverage the disturbance to the wireless signals for monitoring the intruding object. Youssef et al. demonstrate the DfP feasibility and raise the essential challenges in it [26]. Xu et al. propose an active node based method to use the disturbance from the human body to the RF pattern for indoor localization [27]. A following work SCPL is proposed to model the human trajectory through Viterbi algorithm [6]. Compared to our work, these two works use much powerful active tags and unlicensed RF bands. Meanwhile, the location accuracy of SCPL is 1.3m, lower than Twins based tracking scheme. Although the work [8] has an accurate detection rate, the system utilize active tags, which may introduce a high cost in large scale applications. TASA [7] provides the function of tag-free activity sensing or route tracking. TASA also exploit the RSS change for motion detection. TASA, however, still need to involve active tags for achieving reliable sensitivity in the activity sensing. Differ from these previous device-free works, Twins is merely based on passive tags, which is more cost-effective. Meanwhile, Twins proposes a novel detecting method by leveraging the critical state of coupling tags, with much better accuracy.

It has been observed that nearby tags can produce an interference to each other. Weigand and Dobkin conduct experiments over two tag arrays at two planes [14]. The result reported in [14] shows that the successful interrogation rate of tags is severely affected when two parallel arrays are approaching to each other. Later, Chen et al. propose a model for nearby tags to formulate their interference affect [15].

IX. CONCLUSION

In this paper, we propose a novel device-free object tracking scheme, Twins. We contribute to both the theory and practice of a new observed phenomenon, critical state on two adjacent tags. We also design a practical tracking scheme based on our findings. The extensive real experiments demonstrate the effectiveness of our scheme. Our future work includes studying critical state on a single tag, utilizing Twins to track multiple objects, and extending the detection region by refining the tracking algorithms.

REFERENCES

[1] X. Zhu, Q. Li, and G. Chen, "APT: Accurate Outdoor Pedestrian Tracking with Smartphones," in Proceedings of IEEE INFOCOM, 2013.
[2] S. Guha, K. Plarre, D. Lissner, S. Mitra, and B. Krishna, "AutoWitness: Locating and Tracking Stolen Property While Tolerating GPS and Radio Outages," in Proceedings of ACM SenSys, 2010.
[3] L. M. Ni, Y. Liu, Y. C. Lau, and A. Patil, "LANDMARC: Indoor Location Sensing Using Active RFID," *ACM Wireless Networks, (WINET)*, vol. 10, iss. 6, pp. 701-710, 2004.
[4] J. Wang, F. Adib, R. Knepper, D. Katabi, and D. Rus, "RF-Compass: Robot Object Manipulation using RFIDs," in Proceedings of ACM MobiCom, 2013.
[5] J. Maneesilp, C. Wang, H. Wu, and N.-F. Tzeng, "RFID Support for Accurate 3D Localization," *IEEE Transactions on Computers*, vol. 62, iss. 7, pp. 1447-1459, 2013.
[6] C. Xu, B. Firner, R. S. Moore, Y. Zhang, W. Trappe, R. Howard, F. Zhang, and N. An, "Scpl: Indoor Device-free Multi-subject Counting and Localization using Radio Signal Strength," in Proceedings of ACM IPSN, 2013.
[7] D. Zhang, J. Zhou, M. Guo, J. Cao, and T. Li, "TASA: Tag-Free Activity Sensing Using RFID Tag Arrays," *IEEE Transactions on Parallel and Distributed Systems (TPDS)* vol. 22, iss. 4, pp. 558-570, 2011.
[8] Y. Liu, L. Chen, J. Pei, Q. Chen, and Y. Zhao, "Mining Frequent Trajectory Patterns for Activity Monitoring Using Radio Frequency Tag Arrays," in Proceedings of IEEE PerCom, 2007.
[9] J. Han, C. Qian, D. Ma, X. Wang, J. Zhao, P. Zhang, W. Xi, and Z. Jiang, "Twins: Device-free Object Tracking using Passive Tags," *Technical Report*, http://www.cs.utexas.edu/~cqian/Twins.pdf, 2013.
[10] Y. Zhao, Y. Liu, and L. M. Ni, "VIRE: Active RFID-based Localization Using Virtual Reference Elimination," in Proceedings of 34th International Conference on Parallel Processing (ICPP), 2007.
[11] J. Wang and D. Katabi, "Dude, Where's My Card? RFID Positioning That Works with Multipath and Non-Line of Sight," in Proceedings of ACM Sigcomm, 2013.
[12] D. M. Dobkin, *The RF in RFID, Passive UHF RFID in Practice*: Elsevier Inc., 2007.
[13] R. K. Wangsness, *Electromagnetic Fields*: Wiley-VCH, 1986.
[14] S. M. Weigand and D. M. Dobkin, "Multiple RFID Tag Plane Array Effects," in Proceedings of IEEE Antennas and Propagation Society International Symposium, 2006.
[15] X. Chen, F. Lu, and T. T. Ye, "The 'Weak Spots' in Stacked UHF RFID Tags in NFC Applications," in Proceedings of IEEE RFID, 2010.
[16] A. Doucet, N. D. Freitas, and N. J. Gordon, *Sequential Monte Carlo Methods in Practice*: Springer, 2001.
[17] M. K. Pitt and N. Shephard, "Filtering via Simulation: Auxiliary Particle Filters," *Journal of the American Statistical Association*, vol. 94, iss. 446, pp. 590-599, 1999.
[18] P. Mohan, V. N. Padmanabhan, and R. Ramjee, "Nericell: Rich Monitoring of Road and Traffic Conditions Using Mobile Smartphones," in Proceedings of ACM SenSys, 2008.
[19] J. Xiao, K. Wu, Y. Yi, L. Wang, and L. M. Ni, "Passive Device-free Indoor Localization Using Channel State Information," in Proceedings of IEEE International Conference on Distributed Computing Systems (ICDCS), 2013.
[20] J. Liu, B. Priyantha, T. Hart, H. S. Ramos, A. A. F. Loureiro, and Q. Wang, "Energy Efficient GPS Sensing with Cloud Offloading," in Proceedings of ACM SenSys, 2012.
[21] Y. Chen, D. Lymberopoulos, J. Liu, and B. Priyantha, "FM-based Indoor Localization," in Proceedings of ACM Mobisys, 2012.
[22] R. Nandakumar, K. Chintalapudi, and V. Padmanabhan, "Centaur: Locating Devices in an Office Environment," in Proceedings of ACM MobiCom, 2012.
[23] A. Rai, K. K. Chintalapudi, V. N. Padmanabhan, and R. Sen, "Zee: Zero-Effort Crowdsourcing for Indoor Localization," in Proceedings of ACM MobiCom, 2012.
[24] Z. Yang, C. Wu, and Y. Liu, "Locating in Fingerprint Space: Wireless Indoor Localization with Little Human Intervention," in Proceedings of ACM MobiCom, 2012.
[25] J. S. Choi, H. Lee, D. W. Engels, and R. Elmasri, "Passive UHF RFID-Based Localization Using Detection of Tag Interference on Smart Shelf," *IEEE Transactions on Systems, Man, and Cybernetics—Part C: Applications and Reviews*, vol. 42, iss. 2, pp. 268-274, 2012.
[26] M. Youssef, M. Mah, and A. Agrawala, "Challenges: Device-free Passive Localization for Wireless Environments," in Proceedings of ACM MobiCom, 2007.
[27] C. Xu, B. Firner, Y. Zhang, R. Howard, and J. Li, "Improving RF-based Device-free Passive Localization in Cluttered Indoor Environments through Probabilistic Classification Methods," in Proceedings of ACM IPSN, 2012.